\documentstyle[12pt]{article}
\documentstyle[12pt]{article}
\begin{document}
\thispagestyle{empty}
\renewcommand{\thefootnote}{\dagger}

\centerline
{\bf INDUCED GRAVITY FROM THE CONSISTENT STRING THEORY \\}
\vskip 0.3truecm
\centerline
{\bf COUPLED WITH TWO - DIMENSIONAL QUANTUM GRAVITY\\}

\vskip 0.7truecm

\centerline
{\bf I. L. Buchbinder,\footnote{ e-mail adress: josephb@tspi.tomsk.su}}

\bigskip
\centerline{\sl Department of Theoretical Physics.}
\centerline{\sl Tomsk State Pedagogical Institute, Tomsk, 634041, Russia}

\bigskip
\renewcommand{\thefootnote}{\ddagger}
\centerline
{\bf I. L. Shapiro,\footnote { e-mail address:
shapiro@fusion.sci.hiroshima-u.ac.jp}}

\bigskip

\centerline{\sl Department of Mathematical Analysis.}

\centerline{\sl Tomsk State Pedagogical Institute, Tomsk, 634041, Russia}

\centerline{\sl and Department of Physics, Hiroshima University}

\centerline{\sl Higashi - Hiroshima, Hiroshima, 724, Japan}

\bigskip
\centerline
{\bf A. G. Sibiryakov}

\bigskip
\centerline{\sl Department of Quantum Field Theory.}
\centerline{\sl Tomsk State University, Tomsk, 634050,Russia}

\vskip 0.2truecm

\noindent
{\bf Abstract}
The unified theory of string and two-dimensional quantum gravity is considered.
The action for two-dimensional gravity is choosen in a well-known induced form
and thus gravity posesses it's oun nontrivial dynamics even on the classical
level. Three classically equivalent forms of the action of the unified theory
are described, that is $D$, $D+1$ and $D+2$ formulations where  the last one
corresponds to the special kind of nonlinear sigma model on the background of
classical two-dimensional metric. In all cases we find the action of the theory
to be invariant under generalized symmetry transformations, and point out the
arbitrariness of renormalization which reflects this symmetry. The one - loop
counterterms are calculated in a general covariant gauge and also in conformal
gauge. The difference between the counterterms for all classically equivalent
variants of the theory is trivial on mass-shell. Since the different versions
of the theory lead to the one-loop contributions which coincide on mass shell,
the corresponding arbitrariness is related with the only generalized
reparametrizations of the theory. Hence one can formulate the conditions
of Weyl symmetry for any equivalent form of the action. Such a conditions are
written for the $D+2$ formulation and then are reduced to other ones. The
equations for the background interaction fields differ from the ones in
standard
approach and look much more complicated.

\setcounter{page}1
\renewcommand{\thefootnote}{\arabic{footnote}}
\setcounter{footnote}0
\newpage

\section{Introduction} It is well - known that the world sheet of the free
classical string is
flat. The nontrivial two dimensional geometry is an essential quantum
effect which is conditioned by only the Weyl anomaly in the path integral over
the string coordinates [1]. In other words it means that in the framework of
the standard string theory the
two - dimensional gravity is always induced and the classical action for the
two - dimensional gravity is not introduced.

     The string interactions may be discribed within the sigma - model
approach [2 - 4], allowing to construct an effective action for
string exitations (see [5 - 6,8,32,37] for the review). This approach leads
to the string theory in the background fields, corresponding to a condensate
of the massless modes. The background fields obey the equations of motion
following from the principle of Weyl invariance of the quantum theory. In
the theory of bosonic string the background field
of tachyon mode is also considered. Note, that recently there
were an  attempts to take into
account the background fields, which correspond to the massive modes [7 - 11].

     The action of closed boson string in the massless and tachyon
background fields has the form;
$$
S_{str} = \int {d^2}\sigma \sqrt {g} \{ {1\over {2 \alpha'}}
{g^{\mu\nu}} G_{ij}(X)
\partial_{\mu}{X^i} \partial_{\nu}{X^j} + {1\over {\alpha'}}
{\varepsilon^{\mu\nu}\over {\sqrt {g}}} {A_{ij}}(X)
\partial_{\mu} {X^i} \partial_{\nu} {X^j} +
$$
$$
+ B(X)R + T(X)\}  \eqno(1)
$$

     Here $i,j = 1,2,...,D;   {\mu, \nu} = 1,2;   G_{ij}$ - is the background
metric, $A_{ij}$ is the antisymmetric tensor background field, $B(X)$ is the
dilaton field, $T(X)$ is the tachion background field. $R$ is the two -
dimensional curvature, $X^{i}({\sigma})$ are the string coordinates.
     A parameter ${\alpha'}$ may be considered as the parameter of
string loop expansion. From this point of view the Fradkin - Tseytlin
term $BR$ in (1) looks like a
one - loop correction. Therefore in an accord with the previos note
the two-dimensional metric in the
action (1) has no classical dynamics. So even in the presence of the
background fields the classical geometry of the world sheet
remains flat. Nontrivial two - dimensional geometry may arise only
due to the quantum effects.

     Recently there has been a considerable interest to the
two - dimensional quantum gravity. The investigation of the
two - dimensional quantum gravity was inspired by the noncritical string
theory where the well - known
action $\int R\Delta^{- 1}R$ naturally arises. Later the other
two - dimensional actions having no direct links with strings and
with the action $\int R\Delta^{- 1}R$ have been suggested. The analysis
of these theories have been realized both nonperturbatively and
perturbatively [12 - 29, 33 - 36].

     From quantum gravity models point of view the nontrivial
two - dimensional geometry exists even on the classical
level. Since the two - dimensional space is considered as a string
world sheet, the appearance of nontrivial
two - dimensional geometry means that even in a classical theory
string world sheet should be curved from the very beginning.
As a result we face a problem of unification of
two - dimensional gravity theory and string theory.

     The given  paper is just devoted to an attempt to consider such
an unified theory including the string in a background fields
coupled to the
two - dimensional gravity . As a string model we consider a theory with
the action (1). The two - dimensional gravity is described by some
action $S_G$. Thus our aim is to investigate a quantum theory with the
action $S_{str} + S_G$.

The action $S_G$ is taken in the form
$$
S_{g} = \int {d^2}\sigma \sqrt {g} \{ {1\over 2} g^{\mu\nu}
\partial_{\mu}\Phi \partial_{\nu}\Phi + C_{1}R\Phi + V(\Phi)\}  \eqno(2)
$$

     Here $ C_{1}$ is a constant and $V(\Phi)$ is an arbitrary function
of field $\Phi$. The two - dimensional gravity theory with this action
was the object of investigation in the Refs [22 - 29, 33 - 36]. For instance,
in [29, 33, 34]
the one-loop divergences and effective potential have been derived in a linear
covariant gauges with arbitrary gauge parameters.
The features of two-dimensional space lead to some nontrivial properties of the
induced gravity that are related with the gauge dependence[34].

In standard string theory it is accepted to consider the path integral
in two steps [1]. First the integration over string coordinates is
fulfilled and then the ones over the metric $g_{\mu\nu}$. After the first
integration the condition of Weyl invariance is introduced and after this the
integral over metric is transformed to the integral only over
modulus. The basis argument for such a considration is that the
nontrivial two - dimensional geometry is pure induced and can arise only
after integration over string coordinates. In our approach the
nontrivial two - dimensional geometry is introduced on a classical
level. Therefore we will understand the path integration over $X^i$
and $g_{\mu\nu}$ as it is generally accepted in quantum field theory and will
not follow the order of integration. At the same time there are strong reasons
to input the condition of Weyl symmetry at quantum level (see, for example,
[37].
Thus we obtain the effective equations for the background fields which can
differs
from the standard ones due to the nontrivial contributions of the
two-dimensional
quantum metric.

We report here about the one-loop calculations in the theory with action
$S_{str} + S_{g}$ and discuss the gauge
dependence and classical symmetries of the
theory. Firstly we derive the divergent part of the effective action in a
harmonical type gauge with arbitrary
parameter, and then in the more usual conformal gauge where the unified model
of
$D$ - dimensional nonlinear sigma-model coupled with gravity
is reduced to the $D+2$
 - dimensional nonlinear sigma-model with the special form of the background
metric
in $D+2$ - dimensional target space.
In such a formulation the background fields $G_{AB},\;A_{AB},\;{\cal B}$ of
$D+2$ dimensional nonlinear sigma model have the special restricted form (4),
(9.
Therefore we can apply the well known methods and results of the
standard sigma model
approach to our unified model. The main problem here is to check the
equivalence of a
different formulations of the theory on the quantum level.
Since our main purpose is to get the condition of Weyl
invariance for the theory (3),
the formulation (8) is of special interest. One can write down the
corresponding consistency
equations in the usual manner within the $D+2$ dimensional nonlinear
 sigma model without
quantum gravity and then reduce them to the original formulation (3).
The only problem is to
check that the reformulation of the original
theory is equivalent to the reparametrization
transformation of the quantum variables and corresponding background
(interaction) fields.
We solve this problem in a following way.

The difference between two
effective actions, corresponding to different choice of gauge,
is proportional to
the classical equations of motion. Since the change of gauge condition
corresponds
to some change of variables [30] one can consider this arbitrariness as the
consequence of the special extra symmetry, which takes place in the theory.
This symmetry can be viewed as the
manifestation of the standard reparametrization dependence for
$D+2$ - dimensional
theory. Since at the one-loop level these reparametrization does not violate
the
special form of the $D+2$ - dimensional background metric
(as it will be shown below
the block structure of the matrices $G_{AB},\;A_{AB}$ is not violated by
renormalization
even at higher loops), the theory under
consideration is renormalizable at least at the one-loop level, and we can
construct
the beta-functions in a usual manner. Then, taking into account the Weyl
noninvariance of the classial action, we obtain the effective equations for the
background fields. The gauge dependence of the beta functions corresponds to
the usual
reparametrization of the fields, that have usual tensor
(with respect to target space)
form. Therefore the gauge invariance is irrelevant, and we can restrict
ourselves
by the special choice of the gauge, and write the more simple form of
the effective equations.
-1~
The paper is organized as following. In section 2 we start with the detailed
description of the unified model, and find the relation between three different
formulations of the theory, that are i) $D$ - dimensional sigma-model (1)
coupled
with gravity (2), ii)$D+1$ - dimensional sigma-model of special restricted type
coupled with gravity, however there is no special action for gravity in this
case, and
iii) $D+2$ - dimensional nonlinear sigma-model of more restricted form without
quantum gravity. The symmetries of the unified model under consideration are
explored in section 3, where we
apply these symmetries to the investigation of the
renormalization structure of the theory.
Section 4 is devoted to the one-loop calculation in harmonical covariant
bachground
gauge. This calculation is nontrivial because of necessity to preserve both
general
coordinate invariance in $D$  - dimensional target space and on
two-dimensional world
sheet. We solve this problem appliing the background field method and
Schwinger-DeWitt technique (with the use of some calculational modifications,
basically introduced in [23,24,29,33,34]) in the gravity sector
and simultaneously
the normal
coordinate method in target space. All necessary notations
related with the tensor analysis
in target space are introduced in Appendix 1. The calculation in a
harmonical type gauge
with arbitrary gauge parameter is strongly simplified by some
two-dimensional identity
that is  proved in Appendix 2.

In section 5 we calculate the divergent part of the effective action
within the conformal
gauge. The reduction formulas for the transition from $N+1$ dimensional space
to
$N$ dimensional are written in Appendix 3. Sections 6 and 7 contains the main
result of
the present paper that are one-loop renormalization
and the one-loop conditions of
Weyl invariance. The renormalization of the unified model have rather
complicated
structure because
it necessary include the conformal transformation of the two-dimensional
metric (just as
induced gravity (2)) and contains some arbitrariness.
The formulation of the conformal invariance conditions on quantum level for the
quantum gravity theory coupled with sigma model is also nontrivial due to
 the dynamical
origin of the metric.
 We construct these  conditions within the $D+2$ dimensional
formulation where the two dimensional metric can be regarded as the classical
background,
and then reduce the effective equations to other formulations.
As a result we get the new
form of string effective equations which are much more
complicated then the standard ones.
The results are summarized and discussed in section 8.
It is pointed out that the new
effective equations contains fourth powers of derivatives in a dilaton sector
and
second powers in the sector of target space metric. Thus the resulting
effective
geometry corresponds to the general structure of induced
action which we meet in a quantum theory of conformal factor generated by
 conformal trace
anomaly of matter in external gravitational field.

\section{Three forms of the classical action}

 We start with the discussion of some imporant properties of the classical
action $S = S_{str} + S_{g}$
$$
S = \int {d^2}\sigma \sqrt {g} \{
 {1\over {2 \alpha'}} { g^{\mu\nu}} G_{ij}(X)
\partial_{\mu}{X^i} \partial_{\nu}{X^j} + {1\over
{\alpha'}}{\varepsilon^{\mu\nu}\over {\sqrt {g}}} {A_{ij}}(X)
\partial_{\mu} {X^i} \partial_{\nu} {X^j} +
$$
$$
+ B(X)R + T(X,\Phi)+ {1\over 2} g^{\mu\nu}
\partial_{\mu}\Phi \partial_{\nu}\Phi + C_{1}R\Phi\} \eqno(3)
$$

      Note that we have substituted the sum of tachyon term and the
potential $V({\Phi})$ by the more general expression $T(X,\Phi)$.
The action (3) is not Weyl invariant on the classical level,
so the trace of the classical
energy - momentum tensor is not equal to zero. At the same time one
can hope to provide the
invariance under Weyl transformations taking into
account the quantum corrections to
classical action. To fulfil this program we need to derive quantum corrections
to the trace of the Energy - Momentum Tensor with the
accuracy at least to one loop, and take into account the complete
set of symmetries of the
theory. Both points need better understanding of the structure of
 classical action
of the theory. That is why we consider some classically equivalent
formulations of the theory.

The action (3) may be viewed as the $D + 1$ -  dimensional
${\sigma}$ - model action with the metric $G$, antisimmetric tensor $A$
and dilaton $B$ which have the special form in $D + 1$ target space. The new
variables and fields are defined as follows.
$$
{Y^a} = ({X^i},{\Phi})  ; \;\;\; a = 1,...,D + 1.
$$
$$
G_{ab} = diag(G_{ij},{\alpha'}),\;\;   A_{ab} = diag(A_{ij},0),\;\;
{\cal B}({Y^a}) = B({X^i}) + C_{1}{\Phi},   \eqno(4)
$$

Then the action (3) takes the form
$$
S = \int {d^2}\sigma \sqrt {g} \;\{
 {1\over {2 \alpha'}} { g^{\mu\nu}} G_{ab}(Y)
\partial_{\mu}{Y^a} \partial_{\nu}{Y^b} +
$$
$$
{1\over{\alpha'}}{\varepsilon^{\mu\nu}\over {\sqrt {g}}}\; {A_{ab}}(Y)
\partial_{\mu} {Y^a} \partial_{\nu} {Y^b} + {\cal B}(Y)R + T(Y)\} \eqno(5)
$$

The formal difference between (3) and (5) is the choice of notations for the
variables and fields. In this sence (5) may be considered as the compact form
of
(3) and we shall use both forms of the action keeping in mind the
special structure of the background fields (4) of the theory (5). Of
course, it is quite possible to reject the restrictions on the fields (4)
and to consider the theory with the classical action (5). Note that such
theory have been considered in Ref's [31,32]. The only difference is
that we carry
 out the explicit calculations and arise at the nontrivial gravity to the
effective equations.

Further we shall need the equations of motion of the theory, which follows from
the action (4).
For the sake of simplicity we can write only some (most relevant, because all
the
information is preserved) combination
of this equations, corresponding to the equivalent form (5).
$$
T - {\Delta}{\cal B} = 0
$$
$$
- T + {\cal B}_{ab} { g^{\mu\nu}} \partial_{\mu}{Y^a} \partial_{\nu}{Y^b} +
{{\cal B}_a} {H^a_{bc}} {\varepsilon^{\mu\nu}\over {\sqrt {g}}}
\partial_{\mu} {Y^b} \partial_{\nu} {Y^c} + {\alpha'}{{\cal B}_a}{{\cal B}^a}R
+
{\alpha'}{{\cal B}_a}{T^a} = 0             \eqno(6)
$$

Here ${\Delta} = g^{\mu\nu}{\nabla}_{\mu}{\nabla}_{\nu}$, and other
necessary notations are
introduced in Appendix 1. The indexes $a,b,c$ near
${\cal B},\; T$ and $Y$ indicate the
covariant derivatives in target $D + 1$ dimensional space.
(6) can be easily rewritten in
terms of the original formulation (3)with the help of (4).

One can rewrite action (5) in the form of $D+2$ dimensional nonlinear
sigma-model of a more
restricted form with the use of conformal gauge.
The following form of action (3) corresponds to the sigma - model in $D +
2$ - dimensional target space with the classical two - dimensional gravitation
background.
To reduce (5)
into the $D + 2$  dimensional theory we introduce the conformal gauge
for the two-dimensional metric.
$$
g_{\mu\nu}
= e^{-2\rho}{\bar g}_{\mu\nu}                                       \eqno(7)
$$
where ${\bar g}_{\mu\nu}$ is some fixed metric. Then the eq.(5) takes
the form:
$$
S = \int {d^2}\sigma \sqrt {\bar g} \{
 {1\over {2 \alpha'}} {{\bar g}^{\mu\nu}} G_{AB}(Z)
\partial_{\mu}{Z^A} \partial_{\nu}{Z^B} +
$$
$$
+ \frac{\varepsilon^{\mu\nu}}{\alpha'\sqrt{{\bar g}}} {A_{AB}}(Z)
\partial_{\mu} {Z^A} {\partial}_{\nu} {Z^B} + {\cal B} R({\bar g}) + T \}
\eqno(8)
$$
where
$$
Z^{A} = ({\rho},\;{Y^a}) = ({\rho},\;{X^i},\;{\Phi});\;\;\;\;\;\;\;
   A,B = 1,2,...,D + 2.
$$
$$G_{AB} = \left(\matrix{
              0                         &- 2{\alpha'}{\cal B}_b\cr
              - 2{\alpha'}{\cal B}_b    &{G_{ab}}
\cr}\right)
$$
$$
A_{AB} = \left(\matrix{
              0               &0\cr
              0               &{A_{ab}}
\cr}\right)
\eqno(9)
$$

In quantum region the field ${\bar g}_{\mu\nu}$ remains classical
(see section 5) and
 hence the action (8) may be considered as the form of the action (3)
where the only ${D +2}$
dimensional target space coordinates ${Z^A}$ are quantized but the d = 2 metric
${\bar g}_{\mu\nu}$ is not. All the background fields in (8) are rather
restricted by relations
(9) and (4).

\section{Generalized classical symmetries and the renormalization structure}

The renormalization structure of the theory is defined by the form of
the possible counterterms.
We shall write down all  possible kinds of counterterms and explore
the restrictions
which simplify the renormalization. Moreover we study the arbitrariness of
renormalization related with the symmetries of the unified theory. For these
purposes the
(3) formulation of the theory is favourable since there are no extra
restrictions on the
background fields in this form. To find out the form of the possible
counterterms  and
also the relation between them it is useful to start with the
symmetries of the classical
action.

Action (3) is invariant under the general coordinate transformations in the
two-dimensional and also in $D$ dimensional spaces.
Both transformations do not mix the fields $\Phi$ and $X^i$ sectors of
the action (3).
At the same time the existence of the form (5) of the action indicate to
the possibility
of some more general transformations. If we omit the restrictions (4)
then the corresponding
transformations would be the reparametrizations in the $D+1$  dimensional
 space.
However, because of these restrictions we have to apply only the
transformations
which do not violate the block structure of the matrices
$G_{ab}$ and $A_{ab}$ (4). Furthermore the problem may be extended in
a following way.
Let us denote $V = (g_{\mu\nu},\;\Phi,\;X^i)$ the set of quantum fields of the
theory,
and $\lambda = (C_1,\;G_{ij},\;A_{ij},\;B,\;T)$ the complete set of coupling
constants
and background fields. Now we shall look for the transformations $\delta V$ and
$\delta\lambda $ which remain classical action invariant.
$$
\frac{\delta S}{\delta V}\;\delta V +
\frac{\delta S}{\delta\lambda}\;\delta\lambda = 0 \eqno(10)
$$

Let us firstly use the notations (4),(5). Two dimensional metric
have only one degree
of freedom, and therefore it's variation has to depend on only the one
arbitrary parameter
$\rho$. If we put $\delta g_{\mu\nu} = \rho g_{\mu\nu}$ and
$\delta Y^a = \eta^a$ where
$\rho$ and $\eta^a$ are arbitrary functions of the variables $\Phi$ and $X^i$,
then the condition (10) which relate these
quantities with the corresponding variations of the background fields
$\delta G_{ab},\;\delta A_{ab},\;\delta{\cal B}$ and $\delta T$ can
 be rewritten in the form
$$
\frac{1}{\alpha '}\; G_{c(a}\partial_{b)}\eta^c + {\cal B}_{(a}\; \rho_{,b)} +
\frac{1}{2\alpha '}\delta G_{ab} = 0
   \eqno(11)
$$
$$
\frac{1}{\alpha '}\;\eta^a H_{abc} + \delta A_{bc} = 0
\eqno(12)
$$
$$
\eta^a {\cal B}_{a} + \delta{\cal B} = 0
\eqno(13)
$$
$$
\eta^a T_a + \rho T + \delta T = 0
                                      \eqno(14)
$$
Now we rewrite (11) - (14) in the terms of initial fields, taking $\delta\Phi =
\varphi (\Phi,\;X^i)$ and $\delta X^i = \xi^i(\Phi,\;X^i)$. After
application of (4) we
arise at the following form of (11) - (14)
$$
\varphi' + C_1 \rho' = 0                                        \eqno(11a)
$$
$$
\varphi_{,i} + \frac{1}{\alpha '}G_{ij}\xi'^j +  C_1\rho_{,i} + B_i\rho'= 0
\eqno(11b)
$$
$$
\frac{1}{\alpha '}G_{k(i}\xi^k_{,j)} + B_{(i}\rho_{,j)} + \frac{1}{2\alpha
'}\delta G_{ij} = 0
                                                                   \eqno(11c)
$$
$$
\frac{1}{\alpha '}\xi^i H_{ijk} + \frac{1}{\alpha '} \delta A_{ik} = 0
\eqno(12a)
$$
$$
\varphi C_1 + \xi^i B_i + \Phi\delta C_1 + \delta B = 0              \eqno(13a)
$$
$$
\xi^i T_i + T'\varphi + \rho T + \delta T = 0
\eqno(14a)
$$
where prime stands for the derivative with respect to
$\Phi$ and index $i$ after comma
stands for the derivative with respect to $X^i$.
The consistent analysis of the equations (12a), (14a), (13a), (11c),
(11a), (11b) leads
to the following solutions for the functions $\varphi (\Phi,\;X^i)$ and
$\rho(\Phi,\;X^i)$.
$$
\varphi (\Phi,\;X^i) = -\frac{\Phi}{C_1}\delta C_1 - C_1r(x) -
B\frac{\delta C_1}{C_1^2}
+\psi_0
$$
$$
\rho(\Phi,\;X^i)=\frac{\Phi}{C_1^2}\delta C_1 + r(x)                \eqno(15)
$$
Here $r(x)$ is arbitrary function of $X^i$ and $\psi_0$ and $\delta C_1$ are
arbitrary constants. Moreover there still remains all arbitrariness related
with $\xi^i(X)$, however
the dependence $\xi^i$ on $\Phi$ is removed due to (12a).
Furthermore there is additional arbitrariness related with
numbers $C_1$ and $\psi_0$. The variations of the background
fields are defined by (11a) - (114a). Note that the restricted transformations
(11)-(15) do not violate the block structure of the matrices $G_{ab}$ and
$A_{ab}$
(5) and hence one can consider them as some particular form of the
reparametrization
transformations for the model (5) with the background fielda (4).

Now we can start with the analysis of the renormalization structure.
Since it is possible to consider the theory (3) as a ${D +
2}$ dimensional  ${\sigma}$ - model it is clear that the
renormalization structure of this theory have the features common with
the renormalization structure of the ordinary
${\sigma}$ - model (1).
Note that the more general study of the
 reparametrizations in a theory of
${\sigma}$ - model have been carried out in [30].
At the same time one can expect that we will meet here
the nontrivial features of renormalization of
the local version of induced gravity (2).
The perturbative renormalization of the two - dimensional induced quantum
gravity (2)
have been discussed in [23, 29] (the detailed study including account of the
conformal defects
of the dimensional regularization have been recently given in
[36]). As it was shown in [24,25,29] one can
remove all the divergencies in the induced gravity (2) with the
use of some conformal transformation of the two - dimensional metric together
with the
reparametrization of the field $\Phi$.
So the theory (8) is expected to be renormalizable by power counting.
However at this stage
we have no sure that the theory under discussion possess multiplicative
renormalizability,
because it is not clear whether the block structure of (4),(9) survive
under renormalization.

Taking into account the reasons of general covariance and power counting,
we can easily
write down all possible divergent structures of the effective action.

$$
\Gamma_{div} = \int {d^2}\sigma \sqrt {g} \{ {1\over2}A_1(\Phi,\;X)
g^{\mu\nu}\partial_{\mu}\Phi \partial_{\nu}\Phi + A_2(\Phi,\;X)R +
A_3(\Phi,\;X)+
$$
$$
+ {1\over2}A_{4\;ij}(\Phi,\;X)\partial_{\mu}{X^i} \partial_{\nu}{X^j} +
{\varepsilon^{\mu\nu}\over {\sqrt {g}}} A_{5\;ij}(\Phi,\;X)+
$$
$$
A_{6\;i}(\Phi,\;X) {g^{\mu\nu}}\partial_{\mu}\Phi\partial_{\nu}{X^i}
+A_{7\;i}(\Phi,\;X){\varepsilon^{\mu\nu}\over {\sqrt {g}}}\partial_{\mu}
\Phi\partial_{\nu}X^i \}                                           \eqno(16)
$$
where the functions $A_{1,...,7}(\Phi,\;X)$ include the divergent
constants of regularization.

First of all, we note that the divergences of $A_6$ and $A_7$ types are
forbidden by the
simple topological relations. It is easy to see that in any diagramm,
corresponding to these
terms, there is one external line with the derivative of $X^i$ and
simultaneously
external line with the derivative of $\Phi$. The corresponding diagramm
 have to contain
at least one classical vertex of the same type. Since there is lack of
such classical
vertex the divergent structures $A_6$ and $A_7$ are not generated in
any loop order.

The counterterms of $A_6$ and $A_7$ types was the most dangerous ones
for the renormalizability. All
others have the same block structure as classical action and hence they
can be removed by
the renormalization of the metric $g^{\mu\nu}$, field $\Phi$
and also the background fields $G_{ij}(X),\;A_{ij}(X),\;B(X),\;T(X)$.
At that the renormalization of the field $\Phi$ is called to
remove the divergences of $A_2$
type and the
conformal transformation of the metric $g^{\mu\nu}$ have to remove $A_1$.
In this respect
the renormalization of the unified theory (3) is quite similar with
the one in
two - dimensional gravity (2) [25,29]. The peculiarity of our model is the
existence of the extra (with respect to the theory (2)) field $X^i$.
The possible dependence
of $A_4$ and $A_5$ on $\Phi$, as well as more complicated,
if compared with classical theory,
dependence $A_2$ on $\Phi$ probably
leads to that the multiplicative renormalizability require the
dependence on $\Phi$ of  the background metric  $G_{ij}$, antisymmetric field
$A_{ij}$ and  dilaton $B(X)$
to be introduced already on
the classical level, that contradicts to our purposes.

As will be shown below this problem is irrelevant on the one-loop level,
where the theory
under discussion is  multiplicatively renormalizable.
In general, it is possible, that the
same structure of renormalization holds even at higher loops.

Let us now consider the arbitrariness, which takes place in the
renormalization of the
theory (3). It is convenient to be
back to the notations (10). Suppose that the divergences
are removed by the transformation
$$
V^{(0)} = V + Z_V,\;\;\; \lambda^{(0)} =  \lambda + Z_\lambda         \eqno(17)
$$
where $V^{(0)},\;\lambda^{(0)}$ are the bare and  $V,\;\lambda$
 the renormalized quantities.
Let $S_R$ is the renormalized action of the theory in the one-loop
approximation.
$$
S_R = S + \Delta S,  \;\;\;  \Delta S = O(\varepsilon ^{-1})
\eqno(18)
$$
Here $ \Delta S$ is the counterterm which contains the factor of
 ${\varepsilon}^{- 1}$ as well as the
renormalization
constants $Z_V$ and $Z_\lambda$. Then these constants are defined from
the equation
$$
S_R(V,\;\lambda) = S(V^{(0)},\;\lambda^{(0)})                         \eqno(19)
$$

Now
we suppose that the solution  of (19) contains some arbitrariness, that is the
values of $Z_V$ and $Z_\lambda$ can be substituted by $Z_V + \delta Z_V$ and
$Z_\lambda + \delta Z_\lambda$ correspondingly and (19) still holds.
Then, taking into account only the linear
on $\varepsilon ^{-1}$ terms we meet the following equation for $ \delta Z$:
$$
\frac{\delta S}{\delta V}\delta Z_V +
\frac{\delta S}{\delta\lambda}\delta Z_\lambda
= 0                                                              \eqno(20)
$$

The simple comparison
of (10) and (20) shows that the arbitrariness in the renormalization
of the theory is caused
by the generalized symmetry (10) of the classical action and, in
particular, is described by the solution (15).
We are mainly interested in the beta-functions and hence the $\varepsilon
^{-1}$
pole is of special interest. At the same time one can easily check that
the picture is just the same for higher poles.
Thus all the possible kinds
of the renormalization transformations of the theory preserve the
block
structure (4),(9) of the background metric and antisymmetric tensor in $D+2$
dimensional formulation (8) of the theory.
This property
is of crucial importance for this model and it holds at higher loops. At the
same time
it is not for sure that one can provide renormalizability in the framework of
the
original model (3),
because the possible counterterms of $A_{1,2,4}$ types may lead to the
nontrivial mixing
of the field $\Phi$ and string coordinates $X_i$. Below we concentrate
ourselves on the
one-loop case and show that this is not the case at least in this order
of perturbation theory.

\section{Effective action in harmonic gauge}

In the following
two sections we shall carry out the calculation of divergent part of
the one - loop
effective action. To do this we apply the background - field method
simultaneously in two - dimensional and
$D$ - dimensional covariant form. The harmonic - type background
gauge and the calculational method proposed in the papers [23,29] will be used.
The main point of
this method is the special choice of gauge condition which enables us to
investigate the spectrum of the differential operators of the nonstandard form.

Let's make the background shift of the fields $g_{\mu\nu}, X^{i}, {\Phi}$
according to
$$
g_{\mu\nu}\;{\rightarrow}\;{g'_{\mu\nu}} = g_{\mu\nu} + h_{\mu\nu},
$$
$$
h_{\mu\nu} = {\bar h}_{\mu\nu} + {1\over 2}hg_{\mu\nu},
\;\;\;{\bar h}_{\mu\nu}{g^{\mu\nu}} = 0,
$$
$$
{\Phi} \; {\rightarrow} \; {\Phi'} = {\Phi} + {\varphi},
$$
$$
{X^{i}}\; {\rightarrow}\;
{X'^{i}} = {X^{i}} + \frac{1}{\sqrt{\alpha'}}{\pi}^{i}({\xi^{i}})

\eqno(21)
$$
Here the ${\bar h}_{\mu\nu}$
is the traceless part of quantum metric. We divide the
 quantum metric into trace $h$ and  traceless part for convinience.
The last expansion in (21) supposes the standard use of the normal
coordinate method (see,for example, [6] and referencies therein).
In the framework of
the normal coordinate method the quantum field $\pi^i$ is parametrized
by the tangent vector $\xi^i$ in $D$ - dimensional target space.
It is possible to employ the action (5) instead of (3).
Equation (21) corresponds to the following expansion in the action (5)
$$
Y^{a}\; {\rightarrow}\; {Y'^{a}} = {Y^{a}} + {\pi}^{a}({\xi^{a}})
                                                                      \eqno(22)
$$
instead of last two lines in (21). Further we use the expansion (22)
together with action (5)
for brevity. Note that the calculation within the original
formulation
(3) has been performed independently, since the result is just the same we
omit  the corresponding technical details.

The one - loop
effective action is defined by the bilinear (with respect to quantum fields)
part $S^{(2)}$ of the classical action.
Omitting the surface terms and taking into account the
notations and abbreviations of the Appendix 1 we find
$$
S^{(2)} = \int {d^2}\sigma \sqrt {g} \{\; {1\over 2}{\bar h}_{\mu\nu}
[\;{1\over 2}g^{\beta\nu}({\cal B}^\mu \nabla^\alpha -
{\cal B}^\alpha \nabla^\mu)
+{1\over{4\alpha'}}\nabla_\lambda Y^a \nabla^\lambda Y_a
\delta^{\mu\nu,\;\alpha\beta}\;]{\bar h}_{\alpha\beta}+
$$
$$
+{1\over 2}{\bar h}_{\mu\nu}[- {1\over 2} {\cal B}^\mu \nabla^\nu -
{1\over 2} {\cal B}^{\mu\nu}+
{1\over{4\alpha'}}\nabla^\mu Y^a \nabla^\nu Y_a\;]h +
$$
$$
+{1\over 2} h [\;{1\over 2}{\cal B}^\alpha \nabla^\beta
+{1\over{4\alpha'}}\nabla^\alpha Y^a \nabla^\beta Y_a\;] {\bar
h}_{\alpha\beta}+
h[\;\frac{1}{4} {\cal D}_\mu {\cal D}^\mu {\cal B}\;]h +
$$
$$
+{1\over 2}{\bar h}_{\mu\nu}[\;\sqrt{\alpha'}\;
{\cal B}_b {\cal D}^\mu {\cal D}^\nu
+ (2\sqrt{\alpha'}{\cal D}^\mu {\cal B}_b -
\frac{1}{\sqrt{\alpha'}}\nabla^\mu Y_b){\cal D}^\nu +
\sqrt{\alpha'}\;{\cal D}^\mu {\cal D}^\nu {\cal B}_b\;]\eta^b +
$$
$$
+{1\over 2}\eta^a [\;\sqrt{\alpha'}\;{\cal B}_a\nabla^\alpha\nabla^\beta +
\frac{1}{\sqrt{\alpha'}}\nabla^\beta Y_a {\cal D}^\alpha +
\frac{1}{\sqrt{\alpha'}}\;
({\cal D}^\alpha \nabla^\beta Y_a)\;]{\bar h}_{\alpha\beta}+
$$
$$
+{1\over 2}h[\;- \frac{1}{2}\sqrt{\alpha'}\;{\cal B}_b {\cal D}^\mu {\cal
D}_\mu
-\sqrt{\alpha'}\;{\cal D}^\mu {\cal B}_b {\cal D}_\mu +
\frac{1}{2}\sqrt{\alpha'}\;T_b+\frac{1}{2}\sqrt{\alpha'}
\;({\cal D}^\mu {\cal D}_\mu {\cal B}_b)\;]\eta^b+
$$
$$
+{1\over 2}\eta^a [- \frac{1}{2}\sqrt{\alpha'}\;{\cal B}_a
{\cal D}^\mu {\cal D}_\mu + \frac{1}{2}\sqrt{\alpha'}T_a\;]h+
$$
$$
+\eta^a[-G_{ab}{\cal D}^\mu {\cal D}_\mu -
2 H_{acb}\frac{\varepsilon^{\lambda\nu}}
{\sqrt {g}}
\;\nabla^\nu Y^c {\cal D}_\lambda + H_{acd} H_{\;\;eb}^{d}\;\nabla_\nu Y^e
\nabla^\nu Y^c+
$$
$$
+ {\cal K}_{ab} + \alpha'{\cal B}_{ab} + T_{ab}\;]\eta^b
\eqno(23)
$$

Now we have to introduce the gauge fixing term and also take into account the
contributions of ghosts.
The more general form of the gauge fixing action corresponding to
covariant harmonical gauge is following:
$$
S_{gf} = - {1\over 2}\int {d^2}\sigma \sqrt {g}\;{\chi}_{\mu}\;
{G^{\mu\nu}}\;{\chi}_{\nu}                                         \eqno(24)
$$
where ${\chi}_{\mu}$ is the background gauge and $G^{\mu\nu}$ is weight
operator.
$$
{\chi}_{\mu} = {\nabla}_{\nu}{{\-h}_{\mu}^{\nu}} - {\beta}
{\nabla}_{\mu} h - {\gamma}_{a}{\nabla}_{\mu}Y^{a} - E_{\mu}^{\rho\sigma}
h_{\rho\sigma} - F_{{\mu}a}Y^{a},
$$
$$
G^{\mu\nu} = {\tau} {g^{\mu\nu}}                                 \eqno(25)
$$

Here ${\tau},\; {\beta},\; {{\gamma}_a}, \;E_{\mu}^{\rho\sigma},\;
F_{{\mu}a}$ are some arbitrary functions of the dimensionless variables
${\Phi}$ and ${1\over{\alpha'}} X^{i}$.

Let us firstly
make some notes conserning the choice of these functions. For the sake of
simplicity one
can require the bilinear form of the action to be minimal. The last means that
the second
derivatives acting on quantum fields appear only in the combination $\Delta =
\nabla_\nu \nabla^\nu$.
In four dimensional gravity such a condition rigidly restrict the
choice of the gauge
fixing condition
(see, for example, [38]). On the countrary, in our case of two-dimensional
gravity we
can apply the result of Appendix 2 (see also another proof in [33]) for the
particular  case of the flat background metric) and meet the minimal operator
for the arbitrary value of some gauge parameter $\nu$.
At the same time some requirement are imposed and the
functions ${\tau},\; {\beta}$ and ${{\gamma}_a}$ are choosen in a special way
to provide the minimality of the mentioned bilinear form.
$$
{\beta} = 0, \;\; {\gamma}_{a} = -{\nu\over{{\cal B}}},\;\;
{\tau} = {{\cal B}\over{\nu}}                                     \eqno (26)
$$
Furthermore there remains the arbitrariness related with the functions
$E_{\mu}^{\rho\sigma}$ and $F_{{\mu}a}$. The explicit calculations show that
the
divergent part
of the one - loop effective action does not depend on these functions (just as
in the pure gravity theory (2) [29]) and we shall omit these functions in the
intermediate
expressions for compactness, otherwise the formulas look too combersome.

The bilinear part of the total action  $ S + S_{gf} $ have the form
$$
(S + S_{gf})^{(2)} =
\frac{1}{2}\int {d^2}\sigma \sqrt {g}({\bar h}_{\mu\nu},\;h,\;\;\eta^a )
(\hat{H})({\bar h}_{\alpha\beta},\;h,\;\;\eta^b)^T             \eqno(27)
$$
where $T$ denote transposition, and self - adjoint
operator ${\hat H}$ have the form
$$
{\hat H} = {\hat K} {\Box} + {\hat L}^{\lambda} {\nabla}_{\lambda} + {\hat M}
                                                                 \eqno(28)
$$
where ${\hat K}, {\hat L}^{\lambda},{\hat M}$ are c-number operators in the
space
of the fields $({\bar h}_{\rho\sigma},\;h,\; {\xi}^{a})$.
$$
{\hat K} = \left(\matrix{
\frac{{\cal B}}{2\nu} P^{\mu\nu,\;\alpha\beta}& 0 &0\cr
0 & 0 & -\frac{1}{2}{\cal B}_b\sqrt{\alpha'}\cr
0 & -\frac{1}{2}{\cal B}_a\sqrt{\alpha'} & - G_{ab} +
\frac{\nu\alpha'}{{\cal B}} {\cal B}_a {\cal B}_b    \cr} \right) \eqno(29)
$$
where ${\hat P}=P^{\mu\nu,\;\alpha\beta} = \delta^{\mu\nu,\;\alpha\beta} -
\frac{1}{2}g^{\mu\nu}
g^{\alpha\beta}$ is the projector to the traceless states.
$$
{\hat L}^\lambda = \left(\matrix{
\frac{1}{\nu}g^{\beta\nu}{\cal B}^\mu g^{\alpha\lambda}+\frac{1}{2}
(g^{\beta\nu}{\cal B}^\mu g^{\alpha\lambda} -
g^{\beta\nu}{\cal B}^\alpha g^{\mu\lambda})&
-\frac{{\cal B}^\mu}{2}g^{\nu\lambda} & -\frac{1}{\sqrt{\alpha'}}g^{\nu\lambda}
\nabla^\mu Y_b \cr
\frac{{\cal B}^\alpha}{2}g^{\beta\lambda} &
0 & -{\cal B}^\lambda _b\sqrt{\alpha'}\cr
\frac{1}{\sqrt{\alpha'}}g^{\alpha\lambda}
\nabla^\beta Y_a & 0 & L^\lambda_{ab} \cr}
\right)                                                        \eqno(30)
$$
where
$$
L^\lambda_{ab} = \frac{2\nu\alpha'}{{\cal B}}{\cal B}_a {\cal B}^\lambda_b
-\frac{\nu\alpha'}{{\cal B}^2} {\cal B}_a {\cal B}_b {\cal B}^\lambda +
2H_{abc}
\frac{\varepsilon^{\lambda\nu}}{\sqrt{g}} \partial_\nu Y^c
$$
and
$$
{\hat M}= \left(\matrix{
[\frac{1}{4{\alpha'}} \nabla_\lambda Y^c \nabla^\lambda Y_c-
\frac{{\cal B}}{2\nu}R]
P^{\mu\nu,\;\alpha\beta} & -{1\over 2}{\cal B}^{\mu\nu}+\frac{1}{4{\alpha'}}
\nabla^\mu Y^c \nabla^\mu Y_c & 0 \cr
\frac{1}{4{\alpha'}}\nabla^\alpha Y^c \nabla^\beta Y_c & \frac{1}{4}
1{\cal D}_\mu{\cal D}^\mu{\cal B} &  -\frac{\sqrt{\alpha'}}{2}
{\cal D}_\mu{\cal D}^\mu{\cal B}_b \cr
\frac{1}{\sqrt{\alpha'}} {\cal D}^\alpha \nabla^\beta Y_a &
\frac{\sqrt{\alpha'}}{2}T_a & M_{ab}\cr}
\right)                                                             \eqno(31)
$$
where
$$
M_{ab} =
\frac{\nu \alpha'}{{\cal B}}{\cal B}_a \Box{\cal B}_b -
\frac{\nu \alpha'}{{\cal B}^2}\; {\cal B}_a {\cal B}_b^\mu {\cal B}_\mu
- H_{acd} H_{\;eb}^{d}\;\nabla_\nu Y^e \nabla^\nu Y^c+
 {\cal K}_{ab} + \alpha'{\cal B}_{ab}R + \alpha' T_{ab}
$$

The indexes $a,\;b,...$ are lowered by the metric $G_{ab}$
and lifted up with the help of the inverse metric $G^{ab}$.

The ghost action is defined by the usual relation
$$
S_{gh} = \int {d^2}\sigma \sqrt {g}\;{\bar C}^{\alpha}\;M_\alpha^\beta\;C_\beta
$$
$$
{\hat M}_{gh} = M_\alpha^\beta =
\frac{\delta {\chi}_{\alpha}}{\delta h_{\rho\sigma}}
R_{\rho\sigma ,}^{\;\;\;\;\beta} +
\frac{\delta {\chi}_{\alpha}}{\delta \xi^a} R^{a,\beta}         \eqno(32)
$$
where $R_{\rho\sigma ,}^{\;\;\;\;\beta}$ and $R^{a,\beta}$ are
the generators of the general
coordinate transformations of the corresponding fields
$$
R_{\rho\sigma ,}^{\;\;\;\;\beta} = -\delta_\rho^\beta \nabla_\sigma -
\delta_\sigma^\beta \nabla_\rho
$$
$$
R^{a,\beta} = - \nabla^\beta Y^a                                      \eqno(33)
$$
Sibstituting (33),(26) and (25) into (32) we find
$$
{\hat M}_{gh} = M_\alpha^\beta =
-\delta_{\alpha}^\beta \Box - R_{\alpha}^\beta
+ \frac{\nu}{{\cal B}}\;{\cal B}_a\; \nabla^\beta Y_a+
\frac{\nu}{{\cal B}}\nabla^\beta Y_a\; {\cal D}_\alpha {\cal B}_a \eqno(34)
$$

The one - loop divergencies of the effective action are given by
the expression
$$
\Gamma^{(1)} = -{1\over 2} \left.Trln{\hat H} \right|_{div} +
 \left.Trln{\hat M}_{gh} \right|_{div}                                \eqno(35)
$$

Note that the form of the operator ${\hat M}_{gh}$ enables us to
apply the standard
technique of the
divergences calculation. Furthermore, the structure of operator ${\hat H}$
(28) is suitable for
the use of the calculational method proposed in Ref's [23,29] for ``pure''
gravity (2).
Let us start with the
first term in (35). The operator, inverse to ${\hat K}$, (29) have the
form
$$
{\hat K}^{-1} = \left(\matrix{
\frac{2\nu}{{\cal B}} P^{\mu\nu,\;\alpha\beta}& 0 &0\cr
0 & \frac{4}{\alpha'{\hat {\cal B}}^2} - \frac{4\nu}{{\cal B}}
& -\frac{2{\cal B}^b}{\sqrt{\alpha'}\;{\hat {\cal B}}^2}\cr
0 & -\frac{2{\cal B}^a}{\sqrt{\alpha'}\;{\hat {\cal B}^{2}}}  &
 - G^{ab} +
\frac{{\cal B}_a {\cal B}_b}{{\hat {\cal B}}^{2}}   \cr} \right)\eqno(36)
$$
As it have been pointed out above the matrix ${\hat K}^{-1}$
depends on arbitrary gauge
parameter $\nu$. Now we transform, as in [23, 24]
$$
Trln{\hat H} = Trln{\hat K} +  Trln ({\hat 1} \Box + {\hat K}^{-1}
{\hat L}^\lambda
\nabla_\lambda + {\hat K}^{-1}{\hat M})
                                                               \eqno(37)
$$

Note that the procedure of complexification of the operator ${\hat H}$ [26]
allow to understand
the transformation (37) as the change of variables in the
path integral over $V =
(h_{\mu\nu},\;\Phi,\;X^i)$ [29] but doesn't change the
expression for $\Gamma^{(1)}$ and thus
is not necessary. First term in (37) does
not give contribution to the divergences, because of
 it's local strusture and therefore can be omitted.
The second term  in (37) have standard form
$$
{\hat 1}{\Delta} + {\hat h}^{\lambda} {\nabla}_{\lambda} + {\hat D}
$$
where ${\hat 1}$ is the unity matrix in
the corresponding space. Therefore we can apply
the general expression for divergences (see, for example, [44])
$$
\left.Trln{({\hat 1}{\Delta} + {\hat h}^{\lambda} {\nabla}_{\lambda} +
{\hat D})} \right|_{div} =
- \frac{1}{\varepsilon} tr\left( \frac{{\hat 1}}{6}R -
\frac{1}{2}\nabla_\lambda
{\hat h}^{\lambda} - \frac{1}{4}{\hat h}^{\lambda}{\hat h}_{\lambda} + {\hat D}
\right)                                                        \eqno(38)
$$
where $\varepsilon = 2\pi(d-2)$ is the parameter of dimensional regularization.
Now it is possible to derive the divergences, substituting the expressions
(34),(36), (29) - (31) into the formulas (37), (38) and (35).
So, after some tedious algebra, we get [41]
$$
{\Gamma}^{(1)}_{div} = - {1\over{\varepsilon}}
\int {d^2}\sigma \sqrt {g}\; \{[\;{{24 - D}\over 12}R + {{\alpha'}\over{2}}
({\cal D}^{2}{\cal B})R - {{\alpha'}\over{2}}
{{\cal B}^{a}{\cal B}_{ab}{\cal B}^{b}
\over {{\cal B}^{c} {\cal B}_{c}}}\;]R -
$$
$$
- {{\nu}\over{{\cal B}}}({\Delta}{{\cal B}} -T) +
{1\over 2} {{\cal K}^{a}_{a}} + {{\alpha'}\over{2}}{\cal D}^{2}T
+ {1 \over {{\cal B}}^{c} {{\cal B}}_{c}}[\;{{\cal B}}^{a} T_{a} -
{{\alpha'}\over{2}}
{\cal B}^{a} {T}_{ab}{\cal B}^{b}\;] -
$$
$$
- {1 \over {2 {\cal B}^{c} {\cal B}_{c}} }
[ \;{1\over{\alpha'}} {\Delta}{\cal B} +
{1\over{2}}{{\cal B}^b}{\Delta{\cal B}_{b}} - {1\over{2}}
g^{\mu\nu}{{\cal B}}^{a}_{\mu} {{\cal B}}_{a \nu} +
$$
$$
+ {{\cal B}}^{a} {{\cal B}}^{b}(g^{\mu\nu}H_{fbc} H^f_{\;\;ea}\;
{\partial}_{\mu}Y^{e}
{\partial}_{\nu}Y^{c} + {{\cal K}}_{ab})-
2{{\cal B}}^{b} {{\cal B}}_{da}{\varepsilon^{\mu\nu}\over {\sqrt {g}}}
H^d_{\;\;eb}{\partial}_{\mu}Y^{a}{\partial}_{\nu}Y^{e}]\}

\eqno(39)
$$

The equation (39) is written in terms of the $D+1$ dimensional
 formulation (5) of the theory.
Note that we didn't use (4) in a process of calculations, and
therefore (39) can be regarded
as divergences for the general sigma model (5) with quantum gravity.
One can consider (39) as
the more general result if compared with $D$ dimensional form
because the restrictions on
the metric (4) have been omitted.
At the same time (39) can be easily reduced to the
original variables (3), and one can find the
corresponding expression below (43).

The divergent part of the one - loop effective action (39) may be compared with
the divergencies in some theories, studied earlier by different authors.
For example,
the Einstein term in (39) coinsides with the same term which was computed in
Ref.[26] for the linear (free) sigma - model (see also [35,36,42,43]) coupled
to
quantum gravity.
So we
obtain that this counterterm does not depend on ${\nu}$, that is the expected
situation for the anomaly contribution.
In fact we have considered more general case since in the present paper
the nonlinear sigma model was
applied for the description of matter fields, and the dependence of
gauge parameter was introduced, whereas in the mentioned
papers the gauge parameter ${\nu}$ was fixed to be equal to unity.
In the case of $\nu = 0$ and for the trivial background fields we
 have good correlation
in other terms as well.
Note that the calculation of the one-loop contributions of gravity to the
matter
fields sector [45] can not be simply compared with our case because of
the nontrivial mixed sector in the operator (28).

Furthermore, if one put ${\cal B} = C_{1}{\Phi}$ and $H_{abc} = 0$ and so
remove
all the $X$ - dependent terms, then (39) coincide with the expression,
which was derived in [29] for pure gravity (2).
Note also that the ${\nu}$ - dependent terms in (39) are proportional to the
equations of motion (6),  and hence disappear on mass shell.
This fact is in a nice accordance with the well -
known general theorem [39,40]. One can consider this property of (39)
as some tool for
checking the correctness of calculations.

Next feature of (39) corresponds to the separation of quantum
gravity contributions, and it
is not so obvious. In fact all the terms in  (39), which are related
with the contributions of
quantum metric, exhibit the dependence on $\nu$ or contain the factor of
 $({{\cal B}_{a} {{\cal B}}^{a}})^{-1}$.
Thus removing all the ${\nu}$ - dependent terms and also the terms
which have ${{\cal B}}_{a} {{\cal B}}^{a}$ in the denominators, we obtain the
well-known result for the $D$ - dimensional sigma - model (1)
without quantum gravity.

The expression (39) looks like very complicated and the appearance of
$ {{\cal B}}_{a} {{\cal B}}^{a} $ terms in some  denominators looks
like quite strange, but
it is not so. The next section will help us to understand this result in the
framework of $D+2$ dimensional formulation.

\section{Calculation in conformal gauge}

As it was already shown in the section 2 the theory (3) can be presented as the
 $D+2$ dimensional nonlinear sigma model on the background of purely classical
two dimensional metric (8).

Here we shall apply the reduction formulas (8,9) to the
derivation of the one - loop divergencies in the theory (3).
Since the transition
to $D+2$ dimensional formulation is related with the use of conformal gauge,
this
way is equivalent to the the direct calculation within this
gauge. The use of conformal gauge leads to some new problem
concerning the application of dimensional regularization [13,35,36]. In fact
the
consistent approach implies the formulation of the theory in dimension
$d = 2 + {\varepsilon}$. However in such a dimension the action (3) is not
scale - invariant and
hence we face the problem of the possible nonanalyticity of the inverse
metric in target $D + 2$ - dimensional space as well as of some other objects
like curvatures.
The explicit analysis of these objects shows that all of them are analytical
and hence the use
of dimensional regularization is consistent. Note that the account
of finite contributions or the higher loop corrections
 leads to some additional oversubtraction problem [13,35,36] but now we
consider only the one-loop divergences.
2
One can divide the fields ${\rho}$
and ${Y^a}$ in the action (8) into background ${\rho},\; {Y^a}$
and quantum $r,\;{P^a}$ parts according to
$$
{\rho} \; {\rightarrow}\;  {\rho'} = {\rho} + r,
$$
$$
{Y^a}\;{\rightarrow}\;{\ Y'^a} = {Y^a} + {P^a}  \eqno (40)
$$
Taking into account the normal coordinate expansion for the $D + 1$ -
dimensional
sigma-model we arise at the following bilinear part of the action (8)
$$
S_{conf}^{(2)} = \frac{1}{2\alpha'} \int {d^2}\sigma \sqrt{g}
\{ g^{\mu\nu}(\partial_\mu r\;\partial_\mu P^a)
\left(\matrix{
 0 & -{\alpha'}{\cal B}_b\cr
-{\alpha'}{\cal B}_a &  G_{ab} \cr} \right)
\left(\matrix{
\partial_\nu r \cr
\partial_\nu P^b \cr} \right)+
$$
$$
+\frac{1}{\alpha'}A_{ab}\frac{\varepsilon^{\mu\nu}}{\sqrt{g}}
\partial_\mu P^a \partial_\nu P^b+{\cal B}^2 R + e^{-2r}T \}     \eqno(41)
$$
The last expression allow the direct use of standard methods.
Taking into account the gauge ghosts contributions and the
relations between the curvatures
in $\;D + 1\;$ and $\;D + 2\;$
dimensions from Appendix 3 we obtain the following expression for
${\Gamma}^{(1c)}_{div}$. The symbol $c$ denote the use of conformal gauge.

$$
{\Gamma}^{(1c)}_{div} =  - {1\over{\varepsilon}}
\int {d^2}{\sigma} {\sqrt {g}}\;
\{ {{24 - D}\over 12}R +{1\over 2} {{\cal K}^{a}_{a}} -
 {1 \over {2 {\cal B}^{c} {\cal B}_{c}} }
[\;{1\over{2}}{{\cal B}^b}{\Delta{\cal B}_{b}} - {1\over{2}}
g^{\mu\nu}{{\cal B}}^{a}_{\mu} {{\cal B}}_{a \nu} +
$$
$$
+ {{\cal B}}^{a}
{{\cal B}}^{b}(g^{\mu\nu}H_{bcd}H^d_{\;\;ea}\; {\partial}_{\mu}Y^{e}
{\partial}_{\nu}Y^{c} + {{\cal K}}_{ab})-
2{{\cal B}}^{b} {{\cal B}}_{da}{\varepsilon^{\mu\nu}\over {\sqrt {g}}}
H^d_{\;\;eb}\;{\partial}_{\mu}Y^{a}{\partial}_{\nu}Y^{e}]
$$
$$
+ {{\alpha'}\over{2}}{\cal D}^{2}T +\frac{1}{{\cal B}^{c}{\cal B}_{c}}
[{{\cal B}}^{a} T_{a} - {{\alpha'}\over{2}}{\cal B}^{a} {T}_{ab}{\cal B}^{b}]-
\frac{1}{2{\cal B}^{c}{\cal B}_{c}}[\;({\cal D}^{2}{\cal B})
{\cal B}_{ab}\; g^{\mu\nu} {\partial_{\mu}}{Y^a} {\partial_{\nu}} {Y^b}+
$$
$$
+({\cal D}^2 {\cal B}){\cal B}_e H^e_{\;\;ab}\;
{\varepsilon^{\mu\nu}\over {\sqrt {g}}}\;
{\partial}_{\mu}Y^{a}{\partial}_{\nu}Y^{b}
+ \frac{T}{\alpha'} + ({\cal D}^{2}{\cal B})
(- T + \alpha'{T}_{a}{\cal B}^{a})\;]+
$$
$$
+\;{{\cal B}^a{\cal B}_{ab}{\cal B}^b \over {2({\cal B}^c {\cal B}_c)^2}}
\;[- T + {\cal B}_{ab}\;g^{\mu\nu}\; {\partial_{\mu}}{Y^a} {\partial_{\nu}}
{Y^b}
+{\cal B}^a H_{abc} \; {{\varepsilon}^{\mu\nu} \over {\sqrt {g}}}
\;{\partial_{\mu}}{Y^b} {\partial_{\nu}} {Y^c} + {\alpha'}{{\cal B}_a}{T^a}]\}

\eqno(42)
$$
The expression (42) is just the known divergent part of the
effective action for the
ordinary (that is without quantum gravity)
sigma - model [3] in the case of restricted background interaction fields (9).
Thus we see that the
denominators in (39) and (42) reflect the form of the
curvature tensor in $D+2$ dimensional
space with special kind of metric (9) in this space.

And so we observe that the expressions for
 ${\Gamma}^{(1c)}_{div}$ and ${\Gamma}^{(1)}_{div}$ do not coincide.
Nevertheless both expressions are correct and moreover they
verify the correctness of each other. Really,  the difference
${\Gamma}^{(1c)}_{div} - {\Gamma}^{(1)}_{div}$ is the linear combination
of the equations of motion (6) as it have to be [41].
$$
{\Gamma}^{(1c)}_{div} - {\Gamma}^{(1)}_{div} = - {1\over{\varepsilon}}
\int {d^2}{\sigma} {\sqrt {g}} \{ [ { {\nu}\over{\cal B} } +
{1 \over {2 \alpha' {\cal B}^{a} {\cal B}_{a}}}] ({\Delta}{\cal B} - T) +
{1 \over {2{\cal B}^{c} {\cal B}_{c}}}
[ {{\cal B}^{a}{\cal B}_{ab}{\cal B}^{b} \over {{\cal B}^{c} {\cal B}_{c}}} -
({\cal D}^{2} {\cal B}) ]\;\cdot
$$
$$
\;\cdot \; [- T + {\cal B}_{ab}\;g^{\mu\nu}
{\partial_{\mu}}{Y^a} {\partial_{\nu}} {Y^b} +
{{\cal B}_a} H^a_{\;\;bc} \; {{\varepsilon}^{\mu\nu} \over {\sqrt {g}}}\;
{\partial_{\mu}}{Y^b} {\partial_{\nu}} {Y^c} +
{\alpha'} R {{\cal B}_a}{{\cal B}^a} + {\alpha'}{{\cal B}_a}{T^a}]\}
$$

 Therefore one can suppose that
${\Gamma}^{(1c)}_{div}$ and ${\Gamma}^{(1)}_{div}$ can be  converted
into each other after some reparametrization transformation
and that the corresponding arbitrariness of the interaction fields
have the well - known Killing form.
Note that the relation between the divergences of the pure gravity (2) in
covariant
and conformal gauges have been discussed in [27, 34, 41].
Just as in present case the
corresponding two
expressions do not coinside for any choise of the gauge parameter $\nu$ [34].

\section{One loop renormalization and beta functions}

Since our main
purpose is the investigation of the model (3), it is more useful to
discuss renormalization in terms of the original formulation.
Applying the reduction
formulas (4) to the expression for divergent part of effective action (39) we
get
$$
{\Gamma}^{(1)}_{div} = - {1\over{\varepsilon}}
\int {d^2}\sigma \sqrt {g}\; \{{{24 - D}\over 12}R + {{\alpha'}\over{2}}
({{\cal D}}^{2}B)R - {{\alpha'}\over{2}}
\frac{B^i B_{ij} B^j}{E} R -
$$
$$
- {{\nu}\over{C_1 \Phi + B}}({\Delta}B + C_1\Delta \Phi -T) +
{1\over 2} {{\cal K}^{i}_{i}} +
{{\alpha'}\over{2}}({\cal D}^{2}T +\frac{1}{\alpha'}T'')+
$$
$$
+ \frac{1}{E}
[\;B^i T_i + \frac{C_1}{\alpha'} T' - \frac{1}{2\alpha'}(C_1\;\;\alpha' B^i)
\left(\matrix{
 T'' & T'_j\cr
 T'_i& T_{ij}  \cr} \right)
\left(\matrix{
C_1 \cr
\alpha' B^j \cr} \right)\;]-
$$
$$
- {1 \over {2E}} [ {1\over{\alpha'}} ({\Delta}(C_1 \Phi + B) +
{1\over{2}}{{B}^i}({\Delta{B}_{i}}) - {1\over{2}}
g^{\mu\nu}B^{i}_{\mu} B_{i \nu} +
$$
$$
+ B^j B^k ({{\cal K}}_{ij} + H_{lji} H^l_{\;\;mk}g^{\mu\nu}\;
{\partial}_{\mu}X^k
{\partial}_{\nu}X^{m}) - 2B^j B_{ik}\;{\varepsilon^{\mu\nu}\over {\sqrt {g}}}
\;H^i_{\;\;lj} \; {\partial}_{\mu}X^k {\partial}_{\nu}X^l]\}

\eqno(43)
$$
where $E = \frac{C_1^2}{\alpha'} + B^i B_i$ and primes stands
for the derivatives with
respect to $\Phi$.

The analysis of
renormalization is rather nontrivial. First of all, it is natural to ask:
what can we expect?
For instance, if we take into account the results of section 3, the mixing
between $\Phi$ and $X^i$ in kinetic term of
(43) looks quite misterious. On the other hand,  we know that
the restrictions (4),
which are the only once, that allow to distinquish $\Phi$ and $X^i$, do
not affect the one-loop calculations, and hence such a mixing is quite natural.
In fact the mixing really
takes place,
but it does not affect the renormalization of the background fields.
The one-loop  renormalization
of the theory include the conformal transformation of the
two-dimensional metric
$$
g_{0\;\mu\nu}= g_{\mu\nu} + \frac{1}{\varepsilon}\rho (\phi,\; X)g_{\mu\nu}
$$
$$
\rho (\phi,\; X) = 2\left[{{\nu}\over{C_1 \Phi + B}}
- \frac{1}{2 \alpha' E}\right]                                    \eqno(44)
$$
which is needed to remove the mixing terms.
All other divergences are easily removed by the
following renormalization of the background fields
$$
G_{0\;ij}=\mu^\varepsilon\; \{ G_{ij}\; +\; \frac{\alpha'}{\varepsilon}
[\;-\; (K_{ij}\; + \;H^k_{\;\;il}H^l_{\;\;jk}\;) -
$$
$$
-\frac{1}{E}(- \;B_{ki}B^k_j\; + \;B^k B^m K_{kijm}) \;-
\; \frac{B^kB_{ki}B^lB_{li}}{E^2}\;] \}
                                                              \eqno(45)
$$
$$
A_{0\;ij}=\mu^\varepsilon\; \{ A_{ij}\; +\; \frac{\alpha'}{2\varepsilon}\;
[\;{\cal D}_k H^k_{\;\;ij} - \frac{1}{E}(B^k B^m {\cal D}_k H_{mij} - 2B^k B_l-
({\cal D}^2 {\cal B}){\cal B}_k H^k_{\;\;ij}\;] \}                    \eqno(46)
$$
$$
B_0 = \mu^\varepsilon\;\{B+ \frac{1}{\varepsilon} [\;\frac{24 - D}{12} -
\frac{\alpha'}{2} ({\cal D}^{2}B - \frac{B^i B_{ij} B^j}{E})\;] \}  \eqno(47)
$$
$$
T_0 = \mu^\varepsilon\;\{T + \frac{1}{\varepsilon} \;
[\;\frac{1}{2}( \alpha' {{\cal D}}^{2}T
+T'') +
\frac{1}{E} [\;B^i T_i + \frac{C_1}{\alpha'} T'\;] -
$$
$$
-\frac{1}{2\alpha' E}(C_1\;\;\alpha' B^i)
\left(\matrix{
 T'' & T'_j\cr
 T'_i& T_{ij}  \cr} \right)
\left(\matrix{
C_1 \cr
\alpha' B^j \cr} \right) - \frac{T}{2\alpha' E}\;] \}                 \eqno(48)
$$

Thus the theory (3) is renormalizable at one loop. However
the conformal transformation of the
metric $g_{\mu\nu}$ is needed  to remove the divergences in a mixed sector.
Note that some
arbitrariness takes place in the above transformations, just as in pure gravity
(2). One can find
that some divergences which are removed here by conformal transformation of
the metric $ G_{ij}$
can be removed by renormalization of the two dimensional metric
$g_{\mu\nu}$. We fix the transformations with the use of two requirements: i)
all mixed terms have
to be removed and the theory have to be finite after renormalization, and
ii)the loop corrections to the beta functions and to the energy-momentum tensor
must have the factor $\alpha'$, that is $\alpha'$
is still loop parameter
in the theory under consideration even if the gravity is quantized.
Note that this condition is violated in the tachyon equation (48).
In fact the described structure of renormalization is not surprising.
We have already faced it
in the pure induced gravity
[25, 29], and in this respect the theory under discussion is similar
to (2), where some arbitrariness
of renormalization also takes place. On the other hand
the renormalization of the
background interaction fields (45) - (48) is performed just in a
sigma model manner.

Note that the necessity to
renormalize $g_{\mu\nu}$ does not appear if we restrict ourselves
by the model (5) without
special conditions (4). Then it is not needed to preserve the block
structure of the matrices
(4) and we can remove all the divergences by the transformations
of the background interaction fields. At the same time the  transformation of
the
metric $g_{\mu\nu}$ is still
possible in this case and hence some arbitrariness also takes
place.

\section{Effective equations following from the Weyl invariance principle}

As we know from the instructive
example of pure induced gravity [29] the arbitrariness
related with the renormalization of the two dimensional metric
$g_{\mu\nu}$ affect the beta
functions and therefore we face the serious difficulty if
 trying to formulate the consistency conditions on this base.
Furthermore in the framework of original model (3)
it is not clear at once how to formulate the condition of conformal invariance
on quantum level. In fact when we consider the theory
where the metric is not background
but dynamical variable it is not clear whether we can
use the  condition $T_{\mu}^\mu$
together with the standard results
one renormalization of composite operators [50]
as the criterium of conformal invariance of the model.

To solve the problem of conformal invariance one can use the
equivalent formulation (8) of the theory.
It is quite possible since the conformal invariance of the effective action of
the model (8) with restricted
background interaction fields reflect
the conformal invariance of the original model (3)
as well as the equivalent model (5).
This fact enables us to construct the consistency
conditions in the framework of (8) in
standard manner and then, taking into account the
restrictions (4) and (9) one can obtain
the conditions of conformal invariance of the
model (3). Indeed, since we consider
here the one-loop contributions to the effective
action, the given above proof of the quantum one-loop equivalence of the three
formulations is very significant.

It is well known that the introduction
of the tachyon term leads to some difficulty
related with the lack of equivalence between the different approaches to the
 string theory[???].
That is why in this section we restrict
our consideration by the theory without tachyon
and write the effective equations for the only $G,\;A$ and $B$ background
fields.
In the $D+1$ dimensional formulation (5) the effective equations have the form
$$
{\bar {\beta}}^G_{ab} = K_{ab} +
\frac{1}{{\cal B}^2}\;[\;{\cal B}_{ac}{\cal B}_b^c -
{\cal B}^c{\cal B}^d K_{acbd} + ({\cal D}^{2} {\cal B}) {\cal B}_{ab}]+
$$
$$
+ \frac{1}{{\cal B}^4}[\frac{1}{4}({\cal B}^2)_{,a}({\cal B}^2)_{,b} -
\frac{1}{2}({\cal B}^2)_{,c}{\cal B}^c \;{\cal B}_{ab}\;]
- \frac{1}{4} H _a^{\;\;cd} H_{bcd} = 0                          \eqno(49)
$$
$$
{\bar {\beta}}^A_{ab} = {\cal D}_c H^c_{\;\;ab} + \frac{1}{{\cal B}^2}
[- {\cal D}^2 {\cal B}\;{\cal B}^c
H_{cab} + 2 {\cal B}^c {\cal B}_{d[a}H^d_{\;\;b]c}
-{\cal B}^c{\cal B}^d\; {\cal D}_c H_{dab}] +
$$
$$
+\frac{1}{{\cal B}^4}\;
{\cal B}^c{\cal B}_{cd}{\cal B}^d {\cal B}^e H_{eab} = 0           \eqno(50)
$$
$$
{1 \over {\alpha'}} {\bar {\beta}}^B = {1 \over {\alpha'}}
\frac{D-24}{48\pi^2}+
\frac{1}{16\pi^2}[4{\cal B}^2 - K - \frac{1}{{\cal B}^2}
({\cal B}_{ab}{\cal B}^{ab}
+{\cal B}^{a}{\cal B}^{b}K_{ab} + ({\cal D}^{2} {\cal B})^2)-
$$
$$
- \frac{1}{{\cal B}^4}
( \frac{1}{4}G^{ab}{\cal B}^2)_{,a}({\cal B}^2)_{,b} - \frac{1}{4} {\cal B}^c
({\cal B}^2)_{,c} {\cal D}^{2} {\cal B}) + \frac{1}{12} H_{abc}H^{abc}] = 0
                                                                \eqno(51)
$$

The above equations are the conditions of Weyl invariance of the theory (5)
with
one-loop corrections. One can forget about the original
model (3) and start with the
unrestricted theory (5). Then
the equations (49) - (51) show that the contributions of
dynamical metric $g_{\mu\nu}$ are relevant and the
effective equations are essentially
modified if compared with the standard ones [3].
Now we can apply (4) and the formulas of
Appendix 3 and derive the effective equations
which arise in the framework of the theory
(3).
$$
{\hat \beta}^A_{\Phi,a}\; =\; {\hat \beta}^G_{\Phi,a} = 0,
\;\;\;\;\;\;\;\; a=(\Phi,\;i),\;\;\;\;\;\; i = 1,...,D             \eqno(52)
$$
$$
{\hat \beta}^G\;=\;K_{ij}+\frac{1}{\frac{C_1^2}{\alpha'} + B^i B_i}
[\;B_{ik}B^k_j\; + \;B^k B^m K_{kijm}\;+\;{\cal D}^{2}B\;B_{ij}\;]+
$$
$$
+\frac{1}{(\frac{C_1^2}{\alpha'} + B^i B_i)^2} [\;\frac{1}{4}({\bar B}^2)_{,i}
({\bar B}^2)_{,j}\; -\;\frac{1}{2}B^k ({\bar B}^2)_{,k}B_{ij} \;]=0   \eqno(53)
$$
$$
{\hat \beta}^A_{ij} \;=\;{\cal D}_k H^k_{\;\;ij}\;+
\;\frac{1}{\frac{C_1^2}{\alpha'} + B^i B_i}\; [\;-{\cal D}^{2}B\;B_k
H^k_{\;\;ij}
+ 2B^lB_{k[i}H^k_{\;\;j]l}-B^k B^l {\cal D}_k H_{lij}   \;]+
$$

$$
+\frac{1}{(\frac{C_1^2}{\alpha'} + B^i B_i)^2}\;B^k B^l B^m B_{kl} H_{mij}
=0                                                             \eqno(54)
$$
$$
\frac{1}{\alpha'}{\hat \beta}^B\;=\;\frac{1}{\alpha'}
\frac{D-24}{48\pi^2}+\frac{1}{16\pi^2}\; \{\;4(\frac{C_1^2}{\alpha'} + B^i B_i)
-K - \frac{1}{\frac{C_1^2}{\alpha'} + B^i B_i}\;[B^{ij}B_{ij}+B^jB^iK_{ij}+
({\cal D}^{2}B)^2]-
$$
$$
-\frac{1}{(\frac{C_1^2}{\alpha'} + B^i B_i)^2}\;[\; \frac{1}{4}({\bar
B}^2)_{,i}
({\bar B}^2)^{,i} - \frac{1}{4}B^i ({\bar B}^2)_{,i}{\cal D}^{2}B\;] +
\frac{1}{12}H^{ijk}H_{ijk}\;\}
\eqno(55)
$$

The renormalization transformations (45)-(48) and the
effective equations (49) - (55)
are the main results of the present
paper. The explicit form of the effective equations
shows that the account of two-dimensional gravity leads to the nontrivial
contributions of these equations. In spite of that the renormalization
transformation and the beta functions possess
some arbitrariness, the general structure  the
effective equations shows that they can
 not be reduced
to the usual form with the use of reparametrization
transformations. It is clear that these
equations are much more complicated and lead to the more reach
geometry if compared with
the standard approach. In particular, the equations
for $G$ and $B$ contains the second
derivatives of dilaton multiplied on the curvature tensor and also fourth
derivatives of dilaton field. At the same time there
is lack of the terms with second order
in curvature and hence some loop ordering of the
powers of curvatures remains.

\section{Summary and discussion}

We have started to investigate the sigma model coupled with quantum gravity.
The two-dimensional metric in the theory under consideration have the
nontrivial dynamics already on the classical level, and these
additional degrees of
freedom turn out to be relevant in the quantum theory.
We input some nontrivial action for the two-dimensional gravity and
therefore   the classical
action of the theory is necessary Weyl - noninvariant due
to the term $C_1{\Phi}R$
and also to the potential and tachyon terms. In spite of this one can
regard this theory to restore the Weyl invariance on quantum level due to the
radiation corrections.
To construct the interaction of sigma model with quantum gravity
we choose the induced form of the gravity action. It is well
known that it is just
the action that arise within the standard approach after intergation over the
the matter (sigma model) variables [1,2,13] and hence this choice is quite
natural. In this sence it is natural that the resulting beta functions can
contain the extra terms which have a more complicated structures which usually
appear only at higher loops. We did not explore this problem, but the relation
between our approach and the standard one looks quite interesting.

Since the induced gravity action (2) can be expressed in a form of
linear sigma model [27] it is not something surprising that the
unified theory under discussion can be regarded as the nonlinear sigma
model of special restricted type (8) in a higher dimensional target
space.  We have verified the one - loop equivalence of the unified
quantum theory (3) of quantum $d = 2$ gravity coupled to the nonlinear
sigma model in $D$ - dimensional target space and the $D + 2$
dimensional restricted sigma model (8) on the classical two -
dimensional gravitation background.  It turns out that the unified
model posess some extra symmetries which preserve the special form of
the $D + 2$ dimensional metric on quantum level and thus provide the
renormalizability.

The renormalizability of the theory enables us to construct the one -
loop $\beta$ - functions and to obtain the conditions of the Weyl
invariance in a usual way within the $D + 2$ dimensional formulation
(8). Then these conditions can be reduced to the variables of the
original unified theory (3) as well as to the variables of the
intermediate (and more general) theory (5).  The resulting effective
equations are much more complicated then the ones which arise within
the standard approach. So the account of quantum gravity contributions
leads to more complicated geometry in target space. In particular we
observe the appearance of some higher derivative terms in the one loop
effective equations, where the dilaton field insertions have higher
powers of derivatives if compared with the Einstein term. It is
interesting to note that the same occures in the field of quantum
field theory in an external gravitational field, where the anomaly
induced action contains the fourth degrees of conformal factor
(dilaton field) which can be regarded as some necessary additions to
the Hilbert-Einstein action [46-49, 38].  The last analogy shows that
the investigation of the effective theory (49) - (55) may have the
special interest. And so we find that incorporation of the quantum
gravity effects leads to a more reach string induced geometry of the
space time.

The last remarks concern some unsolved problems. First of all it
should be very interesting to formulate the conditions of conformal
invariance of the theory on quantum level in terms of original
formulation (3), (5) and taking into account the generalized
symmetries (11) - (15) in an explicit way.  Second, the complicated
structure of the effective equations (49) - (55) encourage us to look
for the corresponding effective action. It should be extremely
interesting to compare these action with the anomaly induced action
for the conformal factor in $D=4$.  Third possible way to extend our
study is related with the exploration of supersymmetry version of the
unified model, that is especially interesting from technical point of
view, because in this case the symmetries on the wourld sheet and in
target space have to be conformed with each other.

\noindent
\section{Acknowledgments} Authors are very
grateful to Professor H.Kawai, Professor A.A.Tseytlin and Professor
I.V.Tyutin for the helpful and useful discussions.  One of the authors
(I.L.Sh.) wish to thanks Particle Physics Group at Hiroshima
University and Theory Division at KEK (Tsukuba) for kind hospitality
and to Professor Y.Okada for invitation to visit KEK. The work of
I.L.B and I.L.Sh. was supported in part by Grant No RT1000 from The
International Science Foundation and by Russian Foundation for
Fundamental Research, project no. 94-02-03234.

\noindent

\section{Appendix 1}

In this Appendix we introduce the notations
related with the tensor calculus in $N$
dimensional space and also the elements of the normal coordinate expansion.
The indices $i,j,...$ take the values $1...D$
when $N\;=\;D$, and the values $1...D+1$ when
$N\;=\;D+1$,  and the values $1...D+2$ when
$N\;=\;D+2$. The last means that we have to identify
$i,j,...$ from this  Appendix with $a,b,...$
from the main text in $D+1$ dimensional, and with
$A,B,...$ in the $D+2$ dimensional spaces, correspondingly.

All the geometrical $N$ - dimensional objects,
considered below, are constructed on the base
of the metric $G_{ij}(X)$ and antisymmetric field
$A_{ij}(X)$.  Below we will omit the arguments
$X$. The torsion field $H_{ijk}$ is defined as the
derivative of the potential $A_{ij}$
$$
H_{ijk} = 3\partial_{[i}A_{jk]}                                    \eqno(A1.1)
$$
For convinience we need two kinds of the curvature tensors.
The first of them is constructed
on the base of only the metric  $G_{ij}(X)$ and have the form
$$
K^i_{jkl} = \partial_k \Gamma^i_{lj} - \partial_l \Gamma^i_{kj} +
\Gamma^i_{kn}\Gamma^n_{lj} + \Gamma^i_{ln}\Gamma^n_{kj}
\eqno(A1.2)
$$
Here $\Gamma^i_{kj}$ is the symmetric connection in the $N$  dimensional space.
$$
\Gamma^i_{kj} = \frac{1}{2}G^{il}(\partial_k G_{lj} + \partial_j G_{lk} +
\partial_l G_{kj})
\eqno(A1.3)
$$
The nonsymmetric connection  ${\check {\Gamma}}^i_{kj}$ is defined as sum of
the symmetric
one and the torsion field.
$$
{\check {\Gamma}}^i_{kj} = \Gamma^i_{kj} + H^i_{jk}
\eqno(A1.4)
$$
The corresponding curvature with torsion may be reduced to the form
$$
{\check {K}}_{ijkl}
 = K_{ijkl} + {\cal D}_k H_{ilj} - {\cal D}_l H_{ikj}-
H_{ikn} H^n_{\;\;lj} - H_{iln} H^n_{\;\;kj}
\eqno(A1.5)
$$
where ${\cal D}_k$ is covariant derivative built with symmetric connection.

When expanding the action (3) we have used the the normal coordinate method in
$N$
dimensional space. Quantum field $\pi^i$ in the expansion
$$
X^i\;\rightarrow \; X'^i + \pi^i
\eqno(A1.6)
$$
is parametrised by the vector $\eta^i$ that is
tangent one to the geodesic line $\rho^i(s)$
which connect the points $X^i$ and $X'^i$.
$$
\rho^i(0) = X^i,\;\; \rho^i(1) = X'^i,\;\; \eta^i = \frac{d \rho^i}{ds}|_{s=0},

\eqno(A1.7)
$$
and $s$ changes from 0 to 1.

Such a parametrization is essentially nonlinear.
$$
\pi^i(\eta) = \eta^i - \frac{1}{2!} \Gamma^i_{j_1 j_2}(X)\eta^{j_1}\eta^{j_2}
- \frac{1}{3!}{\Gamma}^{i}_{j_1 j_2 j_3}(X)
\eta^{j_1}\eta^{j_2} \eta^{j_3} +.... \eqno(A1.7)
$$
where
$$
{\Gamma}^{i}_{j_1 j_2 ... j_n}(X) =
{\cal D}_{j_1} {\cal D}_{j_2}....{\cal D}_{j_{n-2}}
{\Gamma}^{i}_{j_{n-1} j_n}(X)
$$

Now we define the covariant derivative, acting along $\rho^i(s)$ as
$$
{\cal D}_s = \eta^i_s {\cal D}_i; \;\;\;\;
                   \xi^i_s = \frac{d \rho^i(s)}{ds} \eqno(A1.8)
$$
Then we can expand an arbitrary functional $I(X^i)$ in a following covariant
way
$$
I(X^i) = \sum_{n=0}^{\infty} \frac{1}{n!}{\cal D}_s^n I(\rho(s))|_{s=0}
\eqno(A1.9)
$$
When expanding the sigma-model action (1) into series the following formulas
are useful:
$$
{\cal D}_s G_{ij} = 0,\;\;\;  {\cal D}_s \eta^i_s = 0,\;\;\;
{\cal D}_s \partial_{\mu} \rho^i = {\cal D}_{\mu} \eta^i_s,
$$
$$
{\cal D}_s {\cal D}_s \partial_{\mu}\rho^i = {\cal D}_s{\cal D}_{\mu}\eta^i_s =
K^i_{jkl} \eta^j_s \eta^k_s\partial_{\mu}\rho^l
\eqno(A1.10)
$$
For the sake of convinience it is needed to introduce also the covariant
derivative
${\cal D}_{\mu}$ acting on any tensor $T^{i\nu}$ as
$$
{\cal D}_{\mu}T^{i\nu} = \nabla_{\mu}T^{i\nu} +
{\Gamma}^{i}_{jk}T^{k\nu}\partial_{\mu} X^j

\eqno(A1.11)
$$
and also corresponding derivative with torsion
$$
{\check {{\cal D}}}_{\mu}\eta^i =
{\cal D}_{\mu}\eta^i + \eta^k H^i_{jk}\frac{\varepsilon_{\mu\nu}}{\sqrt {g}}
\nabla^{\nu}X^i

\eqno(A1.12)
$$
Some extra notations which have been used in the main text are
$$
B_i = {\cal D}_i B,\;\;\; B_{\mu} = {\cal D}_{\mu} B,\;\;\;
T_{i\mu}={\cal D}_{\mu}{\cal D}_i T

\eqno(A1.13)
$$

etc.

Moreover we use the following notation for $D+1$ space
$$
{\cal K}_{ab} = (g^{\mu\nu} + {{\varepsilon}^{\mu\nu}  \over {\sqrt {g}}})
({\partial}_{\mu}Y^{a})({\partial}_{\nu}Y^{e}) {\check K}_{dabe}(H + {\Gamma})

\eqno(A1.14)
$$
Let us now use the above notations and write
down the expansion of the sigma model action
in a normal coordinates. Since in the main
text of the paper we restrict ourselves by the
only technical details of the $D+1$ dimensional formulation of the theory,
there are
all reasons to make the same here. To get the expansion of (1) it is quite
enough to
change some characters in the following expression.
$$
S[g_{\mu\nu};\; Y^a +\pi^a(\eta)]=\int {d^2}\sigma \sqrt {g} \{
\;(g^{\mu\nu} G_{ab}
+ \frac{2\varepsilon^{\mu\nu}}{\sqrt {g}}\; A_{ab})\;\partial_\mu Y^a
\partial_\nu Y^b+
2{\cal B}(Y)R + T(Y) \}+
$$
$$
+\int {d^2}\sigma \sqrt {g} \{\;  G_{ab}\;\nabla^\mu Y^a\;{\check {\cal D}}_\mu
\eta^b+
\eta^a {\cal B}_a R+T_a\eta^a \}+
$$
$$
+\int {d^2}\sigma \sqrt {g} \{  G_{ab}\;{\check {\cal D}}^\mu \eta^a \;
{\check {\cal D}}_\mu \eta^b +\eta^a\eta^b {\cal B}_{ab}R +
$$
$$
+(\; g^{\mu\nu} + \frac{\varepsilon^{\mu\nu}}{\sqrt {g}} \;)
\;\partial_\mu Y^a \partial_\nu Y^d\;{\check K}_{abcd}(\Gamma + H)\eta^c\eta^b+
T_{ab}\;\eta^a\eta^b \}
\eqno(A1.15)
$$

The above expression have to be supplemented by
the expansion formulas for metric $g_{\mu\nu}$ and also two-dimensional
curvatures etc.
 This expansion of two-dimensional quantities can be easily
found in a few papers and in a
book [38].

\noindent
\section{Appendix 2}

In this Appendix we derive some useful formula for the second order
differential
operators in the space of symmetric traceless second rank tensors.
This formula reflect the
property of two-dimensional space and can not be
extended into another dimension. The projector
into  traceless states have the form
$$
P_{\mu\nu,\alpha\beta}\;=\;\delta_{\mu\nu,\alpha\beta}\;-
\;\frac{1}{2}g_{\mu\nu}g_{\alpha\beta}

\eqno(A2.1)
$$
where $\delta_{\mu\nu,\alpha\beta} = \frac{1}{2}
(g_{\mu\beta}g_{\alpha\nu}+g_{\mu\alpha}g_{\nu\beta})$

Let us start with the flat two-dimensional space
with the metric $\delta_{ab},\;\;a,b=1,2$
One can easily check that for any symmetric second rank tensor $X_{ab}$ it
holds the
following relation
$$
P^{a'b',ab}S_{ab,cd}P^{cd,c'd'} = 0
$$
$$
S_{ab,cd}=\frac{1}{2}\delta_{ab,cd}X_e^e -\delta_{ac}X_{bd}
\eqno(A2.2)
$$

Now we can use the verbeins $e_{\mu}^a$ which obey the relations
$e_{\mu}^a e_{\nu a} = g_{\mu\nu},\;\; e_{\mu}^a e^{\mu b} = \delta^{ab}$
and rewrite (A2.2) for the surface
with arbitrary metric $g_{\mu\nu}$. In such a way we obtain the following
relation
$$
P^{\mu\nu,\mu '\nu '\alpha\beta}S_{\mu '\nu ',\alpha '\beta '}
P^{\alpha '\beta ',\alpha\beta} = 0
\eqno(A2.3)
$$
When receiving the relation (A2.3) the only symmetry property of the  tensor
$X_{\mu\nu}$
have been used. For instance we can take
the symmetric differential operator $X\nabla_{\rho}
\nabla_{\sigma}$ with arbitrary $X$ and
then (A2.3) takes the form of the differential identity
$$
{\bar h}^{\mu\nu}\;X\;[\;\frac{1}{2}\delta_{\mu\nu,\alpha\beta}\Delta -
g_{\nu\beta}
\nabla_{(\mu}\nabla_{\alpha )}\;]\; {\bar h}^{\alpha\beta}
\eqno(A2.4)
$$
where ${\bar h}^{\mu\nu}$ is the
traceless part of the quantum metric, and $\nabla_{\rho}$
is covariant derivative.

Removing the symmetrization and taking into account the curvature properties in
a  two-dimensional space we finally obtain

$$
{\bar h}^{\mu\nu}\;X\;[\;\frac{1}{2}\delta_{\mu\nu,\alpha\beta}\Delta -
g_{\nu\beta}
\nabla_{\mu}\nabla_{\alpha} - \frac{1}{2}\delta_{\mu\nu,\alpha\beta} R\;]\;
 {\bar h}^{\alpha\beta}                 \eqno(A2.5)
$$

In the flat space the last formula have been derived
in [33] as the consequence of relations between the irreducible projectors in
the space
of the fields  ${\bar h}^{\mu\nu}$. From (A2.5) follows that any differential
local
second order operator in this space is minimal, that leads to some nonusual
features
of two-dimensional gravity [34].

\noindent
\section{Appendix 3}

In this Appendix we write down some useful
formulas which relate $N$ and $N+1$ dimensional
geometrical quantities. These  formulas are
extensively used in the main text of the paper
since we consider $D,\;D+1 $ and $D+2$ dimensional
formulations of the original model (3)
and it is needed to relate them with each other. The indices $i,j,..$ take the
values
$1,...,N$ and the indices $a,b,..$ take
the values $1,...,N+1$. The metric of $N+1$ dimensional
space have the following structure.
$$
g_{ab} = \left(\matrix{ a & v_j\cr
v_i& g_{ij} \cr} \right)                                         \eqno(A3.1)
$$
We restrict ourselves by only the case of
$a\;=\; const$ and suppose that the metric $g_{ij}$
and the vector $v_i$ do not depend on zero'th coordinate, moreover we suppose
that
$v_i$ is gradient, that is $v_i = \partial_i v$. ${\cal D}_i$ denote the
covariant
derivative constructed on the base of $g_{ij}$. Furthermore
$v^j\;=\;g^{ij}v_i,\;\;\;
v_{ij}\;=\;v_{ji}\;=\;{\cal D}_j v_i,\;\;\;v^2\;=\;v_iv^i$.
The inverse  $N+1$ dimensional metric
has the form
$$
g^{ab} = \frac{1}{v^2 -a}\left(\matrix{ -1 & v^j\cr
v^i & (v^2 -a)g^{ij} - v^i v^j   \cr} \right)                    \eqno(A3.2)
$$
It is useful to distinquish the $N+1$ dimensional
geometrical quantities (except the metric)
with the help of hats.
The  $N+1$ dimensional Cristoffel symbols are given by expression
$$
{\hat \Gamma}_{0b}^a \;=\;0 ,\;\;\;\;{\hat \Gamma}_{ij}^0
\;=\;-\frac{v_{ij}}{v^2 -a}
,\;\;\;\;
{\hat \Gamma}_{ij}^k\;=\;\Gamma_{ij}^k + \frac{v^k v_{ij}}{v^2 -a}
\eqno(A3.3)
$$
where $\Gamma_{ij}^k$ is the Cristoffel symbol of
the metric $g_{ij}$ in $N$ dimensional space.
According to (A1.2) we find the expression
for the Ricci tensor in $N+1$ dimensional space.
$$
{\hat K}_{bd} = {\hat K}_{\;\;bad}^a,\;\;\;\;\;\;\;{\hat K}_{0a}=0
$$
$$
{\hat K}_{ij}=K_{ij}\frac{1}{v^2}[-v_{ik}v_j^k - v^k v^l K_{iljk}+ v_{ij}{\cal
D}^2]+
\frac{1}{2v^3}[v v_iv_j - v^k v_k v_{ij}]                         \eqno(A3.4)
$$
where $K_{ij}$  is  Ricci tensor of the metric $g_{ij}$ and
${\cal D}^2 = {\cal D}^i {\cal D}_i$.

\newpage
\begin {thebibliography}{99}

\item {Polyakov A.M., {\sl Phys. Lett.} {\bf 207B} (1981) 207.}

\item {Fradkin E.S., Tseytlin A.A.,{\sl Phys.Lett.} {\bf 158B} (1985) 316.;
  {\sl Nucl.Phys.} {\bf 261B} (1985) 1.}

\item {Callan C., Friedan D., Martinec E., Perry M.,
  {\sl Nucl. Phys.} {\bf 272B} (1985) 593.}

\item {Sen A. {\sl Phys. Rev.} {\bf l32D} (1985) 2102.}

\item {Tseytlin A.A., {\sl Nucl. Phys.} {\bf 294B} (1987) 383.}

\item {Osborn H., {\sl Ann. Phys.} {\bf 200} (1990) 1.}

\item {Hughes J., Liu J., Polchinsky J., {\sl Nucl. Phys.} {\bf 316B} (1989)
15.}

\item {Tseytlin A.A., {\sl Int. Mod. J. Phys.} {\bf 4A} (1989) 4249.}

\item {Labastida J.M.F., Vozmediano M.A.H., {\sl Nucl. Phys.} {\bf 312B} (1989)
308.}

\item {Lee J.C., Ovrut B.A., {\sl Nucl. Phys.} {\bf 336B} (1990) 22.}

\item {Buchbinder I.L., Fradkin E.S., S.M.Lyakhovich, V.D.Pershin,
   {\sl Phys. Lett.} {\bf 304B} (1993) 239.}

\item {Weinberg S. in: General Relativity. ed: S.W.Howking and W.Israel,
   Cambridge. Univ.Press. 1979.}

\item {Kawai H., Ninomia M., {\sl Nucl. Phys.} {\bf 336B} (1990) 115.}

\item {Jack I., Jones D.R.T. {\sl Nucl. Phys.} {\bf 358B} (1991) 695}

\item {Polyakov A.M., {\sl Mod. Phys. Lett.} {\bf 2A} (1987) 893.}

\item {Kniznik V.G., Polyakov A.M., Zamolodchikov A.B.
    {\sl Mod. Phys. Lett.} {\bf 3A} (1988) 819.}

\item {David F., {\sl Mod. Phys. Lett.} {\bf 3A} (1988) 1651.}

\item {Distler J., Kawai H., {\sl Nucl. Phys.} {\bf 321B} (1989) 509.}

\item {Chamseddine A.H.,Reuter M.,{\sl Nucl. Phys.} {\bf 317B} (1989) 757.}

\item {Chamseddine A.H. {\sl Phys.Lett.} {\bf 256B} (1991) 379.}

\item {D'Hoker E. {\sl Mod.Phys.Lett.} {\bf 6A} (1991) 745.}

\item {Ichinose S. -Kyoto,1990.-p.51.-(Prepr./Kyoto Univ.-YITP/K-876).}

\item {Odintsov S.D.,Shapiro I.L., {\sl Class.Quant.Grav.} {\bf 8} (1991) 157.}

\item {Odintsov S.D.,Shapiro I.L. {\sl Phys.Lett.} {\bf 263B} (1991) 183.}

\item {Odintsov S.D.,Shapiro I.L., {\sl Europhys.Lett.} {\bf 15} (1991) 575.}

\item {Mazzitelli F.D.,Mohammedi N. -Triest,1992.-p.24.-(Prepr./ ICTP)}

\item {Russo J.G.,Tseytlin A.A. -Cambridge,1992.-p.18.-(Prepr./ DAMTP-1-1992).}

\item {Kantowski R.,Marzban C.,{\sl Phys. Rev.} {\bf 46D} (1992) 5449.}

\item {Odintsov S.D.,Shapiro I.L.,{\sl Int. Mod. J. Phys.} {\bf 1D}
(1993) 571.}

\item {Tyutin I.V. {\sl Soviet J. Nucl. Phys.} {\bf 35} (1982) 222.}

\item {Banks T.,Lykken J. {\sl Nucl. Phys.} {\bf 331B} (1990) 173.}

\item {Tseytlin A.A., {\sl Int. Mod. J. Phys.} {\bf 5A} (1990) 1833.}

\item {Shapiro I.L. {\sl Sov. J. Phys.} {\bf 35,n6} (1992) 69.}

\item {Banin A.T..,Shapiro I.L. {\sl Phys.Lett.} {\bf 324B} (1994) 286.}

\item {Kawai H., Kitazawa Y., Ninomia M., {\sl Nucl. Phys.} {\bf 393B} (1993)
280.}

\item {Kawai H., Kitazawa Y., Ninomia M., {\sl Nucl. Phys.} {\bf 404B} (1993)
684.}

\item {Green M.B., Schwarz J.H., Witten E., {\bf Superstring Theory}
 (Cambridge University Press, Cambridge, 1987.}

\item {Buchbinder I.L.,Odintsov S.D.,Shapiro I.L.,  {\bf Effective Action in
Quantum
Gravity.} IOP Publishing. Bristol and Philadephia, 1992.}

\item {De Witt B.S., {\sl Phys. Rev.} {\bf D} (1967) .}

\item {Voronov B.L., Lavrov P.M., Tyutin I.V.,
                       {\sl Sov. J. Nucl. Phys.} {\bf 36} (1992) 498.}

\item {Buchbinder I.L.,Shapiro I.L.,Sibiryakov A.G., {\bf On string theory
     interacting with two - dimensional gravity} Preprint ICTP. IC/93/206.
Trieste.
     July, 1993.}

\item {Elizalde E.E., Odintsov S.D.,{\sl Nucl. Phys.} {\bf } (1993) 581}

\item {Elizalde E.E., Naftulin S., Odintsov S.D.,{\sl Phys. Rev.} {\bf D}
(1993) 581;
      {\sl Int. J. Mod. Phys.} {\bf A9} (1994) 933}

\item {De Witt B.S.  in: General Relativity. ed: S.W.Howking and W.Israel,
   Cambridge. Univ.Press. 1979.}

\item {Tanii Y., Kojima S., Sakai N.,{\sl Phys.Lett.}{\bf 322B} (1994) 59}

\item {Reigert R.Y.{\sl Phys.Lett.}{\bf 134B} (1984) 56}

\item {Fradkin E.S.,Tseytlin A.A.{\sl Phys.Lett.}{\bf 134B} (1984) 187}

\item {Buchbinder I.L.,Odintsov S.D.,Shapiro I.L.,{\sl Phys.Lett.}{\bf 162B}
(1985) 92}

\item {Antoniadis I., Mottola E., {\sl Phys.Rev.D,} {\bf 45} (1992) 2013}

\item {Osborn H., {\sl Nucl. Phys.} {\bf 294B} (1987) 595; {\bf 308B} (1988)
629}

\end{thebibliography}

\end{document}